# A bright X-ray transient towards NGC 5128 (Centaurus A)⋆

**H. Steinle, K. Dennerl, and J. Englhauser**

Max-Planck-Institut für extraterrestrische Physik, Giessenbachstraße, 85748 Garching, Germany
email: hcs@mpe.mpg.de

**Abstract.** During five ROSAT HRI observations in July 1995, a bright X-ray source (designated 1RXH J132519.8-430312) was detected $2\rlap{.}'5$ south-west of the nucleus of NGC 5128 (Centaurus A) at the outer regions of the elliptical galaxy. At that time it was the brightest point source in the ROSAT HRI field-of-view. All other ROSAT observations made before and after these observations show no trace of the source suggesting that this is a highly variable object. Chandra observations of Cen A made in September 1999 show a source at the ROSAT position which may be the quiescent counterpart of the transient. It is very probable that the transient is located in NGC 5128. Given the existence of a transient source that close to the direction of NGC 5128, all observations (at least in soft X-rays) of Cen A, with instruments of low spatial resolution, must take the presence and variability of this object into account.

**Key words:** Galaxies: individual: NGC 5128 – Galaxies: individual: Centaurus A – X-rays: galaxies – X-rays: stars – novae, cataclysmic variables – Stars: variables: other

**Table 1.** Journal of all pointed ROSAT observations of NGC 5128. All are publicly available from archival data.

| start (yy-mm-dd) | duration (s) | flux* (erg cm$^{-2}$ s$^{-1}$) | instr. |
|---|---|---|---|
| 90-07-27 | 19 662 | $< 2.4 \cdot 10^{-14}$ | HRI |
| 92-01-26 | 13 682 | $< 6.0 \cdot 10^{-14}$ | PSPC |
| 94-08-10 | 64 751 | $< 3.7 \cdot 10^{-14}$ | HRI |
| 95-07-13 | 4 959 | $(3.4 \pm 0.3) \cdot 10^{-12}$ | HRI |
| 95-07-14 | 5 385 | $(3.6 \pm 0.3) \cdot 10^{-12}$ | HRI |
| 95-07-18 | 5 782 | $(3.2 \pm 0.3) \cdot 10^{-12}$ | HRI |
| 95-07-22 | 4 147 | $(2.3 \pm 0.3) \cdot 10^{-12}$ | HRI |
| 95-07-23 | 4 106 | $(3.3 \pm 0.3) \cdot 10^{-12}$ | HRI |
| 98-02-07 | 17 449 | $< 1.3 \cdot 10^{-14}$ | HRI |
| sum 95 | 24 379 | $(3.1 \pm 0.2) \cdot 10^{-12}$ | HRI |
| sum other HRI | 101 862 | $< 2.2 \cdot 10^{-14}$ | HRI |

*: Fluxes are derived assuming a power-law spectrum with photon index $-1.5$ in the 0.1–2.4 keV ROSAT HRI sensitivity range, and an $N_H$ of $8 \cdot 10^{20}$ cm$^{-2}$. Upper limits are $2\sigma$.

## 1. Introduction

The elliptical galaxy NCG 5128 is the stellar body of the giant double radio source Centaurus A (Cen A). It is one example of the family of elliptical galaxies that have an absorbing band of gas and dust projected across their stellar body, obscuring the nucleus at optical wavelengths. The dust lane is thought to be the remnant of a recent ($10^7 - 10^8$ years ago) merger of a giant elliptical galaxy with a smaller spiral galaxy (Thomson 1992).

Cen A as an active galaxy is usually classified as a FR I type radio galaxy, as a Seyfert 2 object in the optical (Dermer & Gehrels 1995), and as a "misdirected" BL Lac type AGN at higher energies (Morganti et al. 1992). It is one of the best examples of a radio-loud AGN viewed from the side of the jet axis (Graham 1979; Dufour & van den Bergh 1979; Jones et al. 1996).

Its proximity of $< 4$ Mpc (Hui et al. 1993) makes it uniquely observable among such objects, even though its bolometric luminosity is not large by AGN standards. Therefore, NGC 5128 is a very well studied and frequently observed galaxy in all wavelength bands. Its emission is detected from radio to high-energy gamma-rays (Johnson et al. 1997; Israel 1998) and it is often used as one of the first targets when new instruments or telescopes are to be tested (e.g. ESO VLT); even space based telescopes are tested on this active galaxy (e.g. HST (Schreier et al. 1996), Chandra (Kraft et al. 2000)) to demonstrate their resolution.

Variability of Centaurus A is reported in many wavelength regimes. In hard and in soft X-rays, observations of the Cen A region have revealed intensity variability greater than an order of magnitude (Bond et al. 1996; Baity et al. 1981; Turner et al. 1997). Many of the observations referred to in the above publications, however, were made by instruments with a spatial resolution much less than required to resolve the inner parts of Cen A.

We here report the detection of a strong source only $2\rlap{.}'5$ off from the nucleus of Cen A in ROSAT High Resolution Imager (HRI) X-ray data. Our data, obtained in 1995, have been recently re-analyzed, and were compared to archival data of all other ROSAT observations of NGC 5128. It turned out that the source is only present in the 1995 observations, implying strong variability. Recently published Chandra X-ray images of Cen A (on the Chandra X-ray observatory public WWW pages) show a weak source at the ROSAT position.

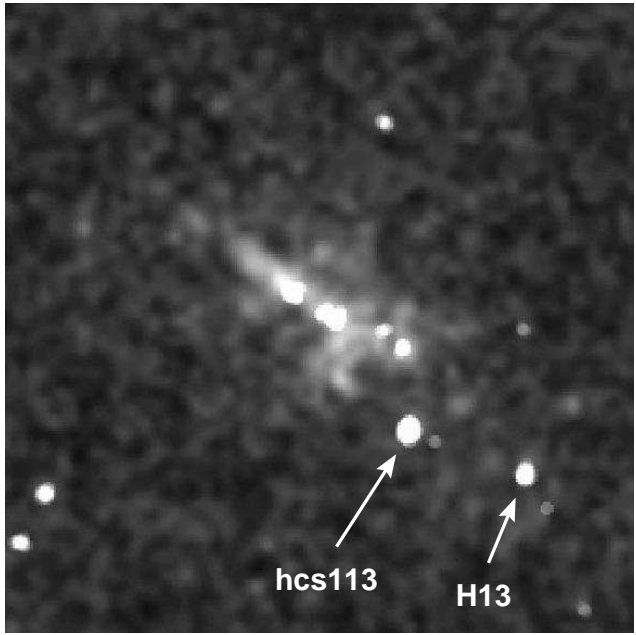

**Fig. 1.** Combined five ROSAT HRI observations from the 1995 multiwavelength campaign which show the strong transient marked as hcs113. For comparison, source H13 from Turner et al. (1997) is also marked. Image size: 12′ x 12′; total exposure 24 379 s; north is up, west to the right.

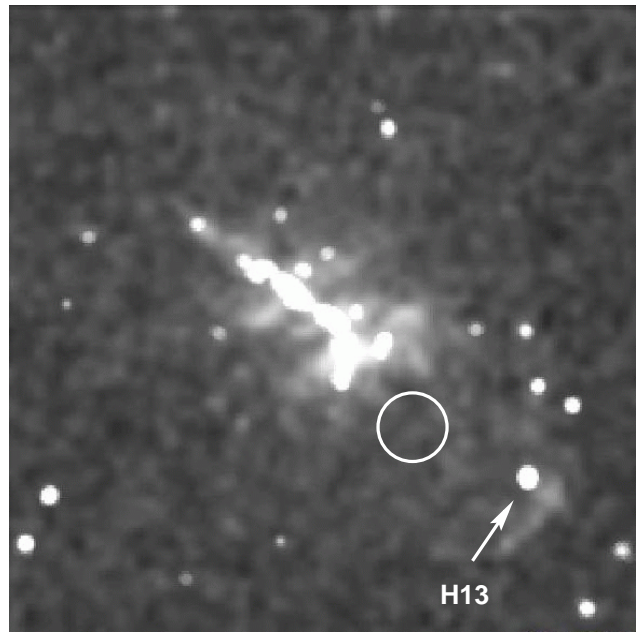

**Fig. 2.** The same region as in Fig. 1. Combined ROSAT HRI observations from 1990, 1994, and 1998 showing no source at the position of the transient which is indicated by the circle. The reference object H13 is marked again. Image size: 12′ x 12′; total exposure 101 862 s; north is up, west to the right.

One important motivation for this *Letter* is to make the existence of a transient source so close to the direction of NGC 5128 known to all observers, inasmuch as all observations of variability (at least in soft X-rays) of Cen A, with instruments of low spatial resolution, must take the presence of this object into account.

## 2. Observations

In July 1995 a multiwavelength campaign took place to observe Cen A (NGC 5128) from radio to gamma-rays. One of the goals of this campaign was to look for correlated variability in different wavelength bands. Over a time interval of 14 days simultaneous measurements with various instruments were made (Steinle et al. 1999). The soft X-ray regime (0.1–2.4 keV) was observed five times with the ROSAT HRI during this time interval. The exposure times were in the order of 5 000 s each. Before and after these 1995 observations four other observations of Cen A were made with ROSAT (three with the HRI and one with the Position Sensitive Proportional Counter (PSPC)): 1990 (HRI, principal investigator (PI) E. Schreier), 1992 (PSPC, PI E. Schreier), 1994 (HRI, PI S. Döbereiner), and 1998 (HRI, PI S. Wagner). All other observations had substantially longer exposure times than the 1995 observations (see Table 1).

## 3. The transient X-ray source 1RXH J132519.8-430312

During the five ROSAT HRI observations in 1995, spanning a time of 10 days (see Table 1), in each observation a bright X-ray source was detected 2′.5 south-west of the nucleus of NGC 5128 at the outer regions of the elliptical galaxy. The coordinates derived from the ROSAT images are R.A. = $13^h\ 25^m\ 19\rlap{.}''8$ and Dec = $-43° \ 03' \ 12''$ (J2000; uncertainty 5″), and the source has been assigned the ROSAT catalogue name 1RXH J132519.8-430312. Throughout the paper, however, we will refer to the transient as hcs113 as it is number 113 in a list of ROSAT HRI / Chandra sources which will be published in a future paper.

In all five observations hcs113 was the brightest point source in the HRI field-of-view with an average count rate of $(0.033 \pm 0.003)$ counts sec$^{-1}$, which is about a factor of 4 brighter than any other point source and about 30 % of the flux of the combined Cen A nucleus and jet sources. In Fig. 1 we show the sum of all five 1995 observations where hcs113 and the object H13 are marked. H13 is a source from Turner et al. (1997)[1] and is used for comparison, as it is a constant source in all ROSAT observations at a count rate of $(0.008 \pm 0.002)$ counts sec$^{-1}$. H13 is probably a 14th mag M star (Feigelson et al. 1981) or a distant early type galaxy (Wagner et al. 1996).

All other ROSAT observations from the years 1990, 1992, 1994, and 1998 show no trace of a source at the position of hcs113, even if combined. In Fig. 2, which is the sum of all ROSAT HRI observations without the 1995 multi-wavelength data, with a total exposure of 101 862 s, the cuts are set as low as possible to detect any object at the transient's position, but no object is detected. The derived $(2\,\sigma)$ upper limit for a detection at the position in question is 0.0003 counts sec$^{-1}$. Due to the long exposure time and the low cuts, a large number of additional sources show up in this image, some of which are visible with different intensity or not at all in Fig. 1. These are candidates for other variable sources (which will be investigated in a future paper), but hcs113 is by far the object with the largest variation.

### 3.1. Simultaneous observations at other wavelengths

As the ROSAT observations in July 1995 were part of an extensive multiwavelength campaign to observe Cen A from radio to gamma-rays, simultaneous observations at other wavelengths exist (Steinle et al. 1999). For the measurements with sufficient spatial resolution (radio, optical, soft X-rays) the observations and analysis concentrated on the nucleus of the active galaxy (this is the reason why the transient went unnoticed until recently). The hard X-ray and gamma-ray results obtained with less spatial resolution (BATSE, OSSE, COMPTEL, and EGRET; all instruments on board the Compton Gamma Ray Observatory (CGRO)) were attributed to the nucleus by the assumption that it is the major source for the high-energy emission.

Besides ROSAT, only the optical monitoring with the ESO 2.2 m telescope at La Silla (Chile), had the position of hcs113 in the field-of-view and had enough spatial resolution. No candidate object in the ROSAT error box down to a limiting magnitude of $\sim 18$ mag is visible in the B and V exposures which were obtined during the same time interval in 1995 (see for example Fig. 3).

### 3.2. Time variability

As listed in Table 1, the measured flux from hcs113 is constant over the 10-day observation period in 1995. The only significant deviation occurs in the observation of July 22, where the flux drops by 30 %. All other ROSAT HRI observations made in 1990, 1992, 1994, and 1998 show no trace of the source. Turner et al. (1997), who analyzed the 1990 ROSAT HRI observation in detail, did not detect any source at the transient's position, nor did they find any object in the ASCA and EINSTEIN data they used for comparison, which would be consistent with the position of hcs113.

---

[1] The paper of Turner et al. (1997) lists all ROSAT HRI count rates about 3 orders of magnitude too high, and Figure 4 of that paper shows the light-curve probably in units erg cm$^{-2}$ s$^{-1}$ and not in photons cm$^{-2}$ s$^{-1}$.

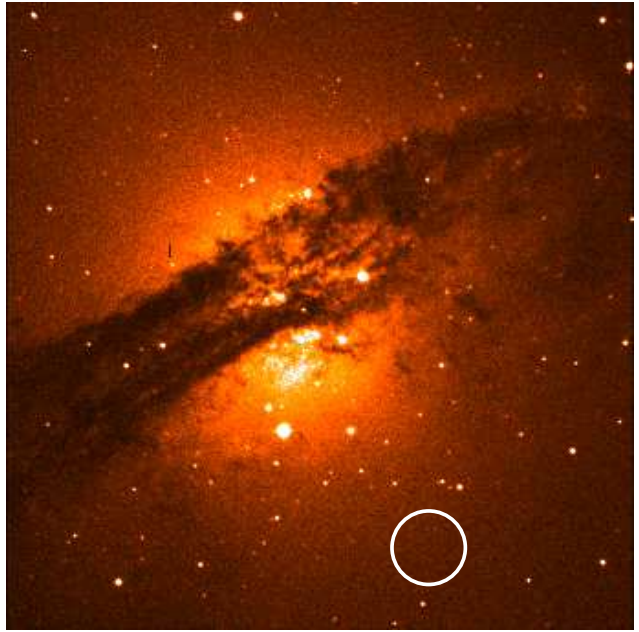

**Fig. 3.** Simultaneous July 1995 optical (B) observation of NGC 5128 made at the ESO 2.2 m telescope at La Silla, Chile. No optical counterpart to the strong X-ray source (position at the center of the circle) is visible. Image size: $7'$ x $7'$ (different from Fig. 1 and Fig. 2 !); north is up, west to the right.

The BATSE instrument on board CGRO has monitored Cen A continuously since its launch in April 1991 in the energy band 20–100 keV, with a very coarse (few degrees) spatial resolution (Wheaton et al. 1996) giving a long baseline for monitoring flux variations. Around the time of the 1995 observations, an increase of the flux is present in the data, but such variations are very common for this source (see Fig. 1 in Steinle et al. (1999)), and similar flux increases occurred during some of the other ROSAT observations, when hcs113 was not detected.

Recently released public images of one of the first Chandra observations, which was imaging the Cen A region with unprecedented arc-second resolution in the energy range 0.1–10 keV, show a weak source at the position of hcs113. This is very probably the transient source either in its quiescent state or in a new outburst phase. However, hcs113 is not one of the highly variable or transient sources listed in the recent paper by Kraft et al. (2000).

*3.3. Spectral information*

As the ROSAT HRI has no energy resolution, no spectral information is available from those observations. Unfortunately the only PSPC observation and the data from the ROSAT All-Sky Survey, which would have had spectral information, were made when the transient was not active. Therefore only limited indirect information can be derived from the simultaneous measurements with the BATSE and OSSE instruments onboard CGRO which observed Cen A in the adjacent higher energy bands.

From the BATSE Cen A monitoring data, no strong enhancement in the 20–100 keV flux is probable (see above). The spectral photon index derived from the data during the 1995 campaign is between $-1.5$ and $-1.7$. A consistent spectral index of $-1.6$ is derived from the OSSE data between 10 keV and 100 keV. These spectral index values are in agreement with the index of $-1.5$ measured by Baity et al. (1981). The conclusion is that the emission of the transient is mainly at soft X-rays.

## 4. Summary and conclusions

During five ROSAT HRI observations in July 1995, a bright X-ray transient source 1RXH J132519.8-430312 (= hcs113) was detected $2\rlap{.}'5$ south-west of the nucleus of NGC 5128. When compared with all other ROSAT observations of Cen A made in 1990, 1992, 1994, and 1998 it turns out, that the source is only present during the 10 day period of the 1995 observations and no trace of it can be detected down to 1 % of the 1995 flux level in the other (deep) observations. Chandra observations of Cen A made in September 1999, however, show a source at the ROSAT position, which may be either the persistent conterpart of hcs113, or the transient at a recent active state.

If at 3.0 Mpc (the distance of Cen A), the luminosity of hcs113 in the energy band 0.1–2.4 keV is $3 \cdot 10^{39}$ erg s$^{-1}$, assuming a power-law spectrum with photon index $-1.5$, $N_H = 8 \cdot 10^{20}$ cm$^{-2}$, and no additional intrinsic absorption in NGC 5128. This is above the Eddington luminosity of most X-ray binaries, but still within the observed range of luminosities for such objects. Therefore it cannot be ruled out that the transient is located in NGC 5128. On the other hand, if hcs113 is a Galactic object, then its soft X-ray luminosity would be less than $3 \cdot 10^{34}$ erg s$^{-1}$ (assumed maximum distance 10 kpc). This is low for a typical neutron star transient, but is high for a typical cataclysmic variable. If it is closer than 10 kpc, its luminosity may be consistent with that of typical cataclysmic variables, whose optical counterparts are usually brighter than 19 or 20 mag, which then should be detectable in (deep) optical images. As the emission of the object seems to be mainly in the soft X-ray range, a distant AGN or Blazar is ruled out. A preliminary search for counterparts of hcs113 has been carried out using the NED (NASA/IPAC Extragalactic Database), the GSC (Hubble Guide Star Catalogue) and SIMBAD without any obvious object being found. It is, however, not the intent of this *Letter* to investigate the nature of this object in more detail.

If hcs113 is indeed an X-ray binary or other variable object, it would be very probable that the observed outburst was not a single event. Therefore, all Cen A observations in the past with low resolution X-ray instruments sensitive in soft X-rays (several keV), may have attributed a variation in the X-ray flux of the transient source to a variation in the flux from the nucleus (and/or jet) of Cen A.

Regardless of the nature and distance of the object, the fact that it is separated by only $2\rlap{.}'5$ from the nucleus of Cen A, poses a strong problem for all spatially unresolved (on this scale) soft X-ray observations. Further observations and the interpretation of previous results (like the Cen A light-curves shown e.g. by Bond et al. (1996) and Turner et al. (1997)) as well as the interpretation of the results of current Cen A monitoring programs in X-rays, as e.g. carried out by RXTE ASM and BATSE (CGRO) have to take the existence of the transient into account.

*Acknowledgements.* This research has made use of the SIMBAD database, operated at CDS, Strasbourg, France, the NASA/IPAC Extragalactic Database (NED), which is operated by the Jet Propulsion Laboratory, Caltech, under contract with NASA, and the ROSAT Data Archive of the Max-Planck-Institut für extraterrestrische Physik, Garching, Germany. We thank the anomymous referee for the helpful comments.

## References

Baity W. A., Rothschild R. E., Lingenfelter R. E., et al., 1981, ApJ 244, 429
Bond I. A., Ballet J., Denis M., et al., 1996, A&A 307, 708
Dermer C. D., Gehrels N., 1995, ApJ 447, 103
Dufour R. J., van den Bergh S., 1979, AJ 84, 284
Feigelson E. D., Schreier E. J., Devaille J. P., et al., 1981, ApJ 251, 31
Graham J. A., 1979, ApJ 232, 60
Hui X., Ford H. C., Ciardullo R., Jacobi G. H., 1993, ApJ 414, 463
Israel F.P., 1998, A&AR 8, 237
Johnson W.N., Zdziarski A.A., Madejski G.M., et al.: 1997, Seyferts and Radio Galaxies. In: 4th Compton Symposium, Dermer C.D., Strickman M.S., Kurfess J.D. (eds.), AIP Conf. Proc. 410, 283
Jones D.L., Tingay S.J., Murphy D.W., et al., 1996, ApJ 466, L63
Kraft R.P., Forman W., Jones C., 2000, ApJ 531, L9
Morganti R., Fosbury R. A. E., Hook R. N., Robinson A. Tsvetanov Z., 1992, MNRAS 256, 1p
Schreier E.J., Capetti A., Macchetto F., Sparks W.B., 1996, ApJ 459, 535
Steinle H., Bonnell J., Kinzer R.L., et al., 1999, Adv. Space Res. 23, 911
Turner T.J., George I.M., Mushotzky R.F., Nandra K., 1997, ApJ 475, 118
Thomson R.C., 1992, MNRAS 257, 689
Wagner S., Döbereiner S., Junkes N.: 1996, X-ray properties of Centaurus A (NGC 5128). In: Röntgenstrahlung from the Universe, Zimmermann H.-U., Trümper J., Yorke H. (eds.), MPE Report (Garching) 263, 521

Wheaton Wm.A., Ling J.C., Mahoney W.A., et al., 1996, A&AS 120, 545